\documentstyle[prb,aps]{revtex}

\begin{document}

\draft

\twocolumn[\hsize\textwidth\columnwidth\hsize\csname
@twocolumnfalse\endcsname
 
\title{ Intermediate Valence Model for the Colossal 
Magnetoresistance in Tl$_{2}$Mn$_{2}$O$_{7}$ }

\author {C.I. Ventura and B. Alascio~$^*$}

\address {Centro At\'omico Bariloche, 8400-Bariloche, Argentina.}

\maketitle

\begin{abstract}
{\small            
The colossal magnetoresistance exhibited by Tl$_{2}$Mn$_{2}$O$_{7}$ is
an interesting phenomenon, as it is very similar to that found in
perovskite manganese oxides although the compound differs both in its 
crystalline structure and electronic properties from the manganites.
At the same time, other pyrochlore compounds, though sharing the same 
structure with Tl$_{2}$Mn$_{2}$O$_{7}$, 
do not exhibit the strong coupling between magnetism and transport
properties found  in this material. Mostly due to the absence of
evidence for significant doping into the Mn-O sublattice, and the 
tendency of Tl to form conduction bands, the traditional double
exchange mechanism mentioned in connection with manganites does not
seem suitable to explain the experimental results in this case. 
We propose  
a model for Tl$_{2}$Mn$_{2}$O$_{7}$ consisting of a lattice of 
intermediate valence ions
fluctuating between two magnetic configurations,
representing Mn-$3d$ orbitals, hybridized with a conduction band, 
which we associate with Tl. This model had been proposed originally for
the analysis of intermediate valence Tm compounds. With a simplified 
treatment of the model we obtain the electronic structure and
transport properties of Tl$_{2}$Mn$_{2}$O$_{7}$, with good qualitative
agreement to experiments. The presence of a hybridization gap in 
the density of states seems important to understand the reported Hall
data.}
  
\end{abstract}


\vskip2pc] \narrowtext



\section{INTRODUCTION}
\label{sec:intro}

	Perovskite manganese oxides like La$_{1-x}$M$_{x}$MnO$_{3}$
(M=Ca,Sr,Ba) have been the subject of study for many years \cite{1}, 
and a strong correlation between transport properties and magnetism
 was recognized. Shortly after the
discovery of these materials, the double exchange mechanism was
proposed \cite{2} to describe the interactions between the Mn ions.
Due to the divalent substitution for La$^{3+}$, itinerant $e_{g}$
holes are doped into the antiferromagnetic insulating parent compound
LaMnO$_{3}$, and Mn ions appear in a mixed valence state.  
Double exchange occurs between heterovalent Mn pairs 
(Mn$^{3+}$-Mn$^{4+}$) by simultaneous transfer of an electron from 
Mn$^{3+}$ to O$^{2-}$ and O$^{2-}$ to Mn$^{4+}$. As this transfer
occurs necessarily conserving the spin orientation, an effective
ferromagnetic coupling appears due to the intra atomic Hund's rule
requirement \cite{2}. Clearly, in this framework the carrier hopping
is intrinsically coupled to and enhanced by the mutual alignment of the
two magnetic moments. As a result, resistivity will depend on the spin
disorder and is expected to display pronounced features at the
ferromagnetic ordering transition temperature (T$_{c}$). 
Application of a magnetic field, which favours alignment of the local spins,
will produce a decrease of the resistivity.

	The interest in the study of perovskite manganese oxides
was renewed recently, as colossal
magnetoresistance (denoted CMR hereafter) was observed near the 
ferromagnetic ordering
temperature \cite{3}. At present, there is strong debate on whether 
the double exchange mechanism and theoretical models based solely 
on it are able to account quantitatively  for the observed 
transport and magnetic properties of 
the manganese perovskites. It has been proposed that other ingredients
such as disorder \cite{4}, Jahn Teller distorsions \cite{4'}, charge
ordering \cite{dagotto}, etc., could play an important role and 
should be taken into account.

	In 1996 colossal magnetoresistance was reported for the
non-perovskite Tl$_{2}$Mn$_{2}$O$_{7}$ compound \cite{5,6,7}. The
material undergoes a ferromagnetic transition 
 with T$_{c}$$\sim$ 140 K \cite{5,6,7,5'}. Below the ordering
temperature, the compound is ferromagnetic and metallic, 
whereas above T$_{c}$ it is paramagnetic. The 
magnetoresistance maximum around the ferromagnetic ordering
temperature is similar to that obtained for the manganese perovskites.
The sharp decrease in resistivity with the development of spontaneous 
magnetization below T$_{c}$ and the magnetoresistance with field
induced magnetization above T$_{c}$ indicate a strong
coupling between transport and magnetism in Tl$_{2}$Mn$_{2}$O$_{7}$, 
similar to that in perovskite manganese oxides \cite{5,6,7,5'}. 
Nevertheless, important differences arise regarding crystalline
structure and electronic properties.
In fact Tl$_{2}$Mn$_{2}$O$_{7}$ is characterized by a ``pyrochlore'' 
A$_{2}$B$_{2}$O$_{7}$ structure \cite{7'},  
consisting of strongly distorted AO$_{8}$ cubes and slightly distorted 
 BO$_{6}$ octahedra. Each metal atom (A or B) forms a
three-dimensional network of corner-sharing tetrahedra \cite{sub0}.
The interpenetrating sublattices in 
Tl$_{2}$Mn$_{2}$O$_{7}$ are respectively given by Tl$_{2}$O and 
Mn$_{2}$O$_{6}$ \cite{5,6,bsub}.  
 The  
tetrahedral Mn$_{2}$O$_{6}$ network differentiates the pyrochlores
from the perovskite ABO$_{3}$ structure, with a cubic MnO$_{6}$
network. There are also significant differences in the bond length and
angle between Mn and O in the MnO$_{6}$ octahedra, which in 
Tl$_{2}$Mn$_{2}$O$_{7}$ are both smaller than for CMR 
perovskites \cite{7,5',5} as well as almost temperature independent,
thereby indicating negligible correlations between spin and lattice
\cite{5'}. 
Considering electronic properties, an important difference with the hole doped
perovskites comes from Hall experiment data \cite{5}, indicating 
a very small number of electron-like carriers: $\sim$~0.001-0.005 conduction 
electrons per formula unit. 
The authors of
Refs.[7,9] mention that such Hall data could result from a small number
of carriers in the Tl-$6s$ band. Assuming
Tl$_{2-x}^{3+}$Tl$_{x}^{2+}$Mn$_{2-x}^{4+}$Mn$_{x}^{5+}$O$_{7}$ with 
x$\sim$ 0.005 the data could be accounted for \cite{7}. 
This would seem to indicate
a very small doping into the Mn$^{4+}$ state \cite{5,7}, in contrast
to manganese perovskites where CMR is obtained around 
30 \% of hole doping. It is important to take notice also of the 
different origin of the doping. Recent electronic structure calculations 
for Tl$_{2}$Mn$_{2}$O$_{7}$ \cite{bsub} indicate that the bands 
inmediately below the Fermi 
level correspond mainly to Mn-$t_{2g}$ states, above the Fermi level appear  
Tl-$6s$ bands, while O states are mixed with these around 
the Fermi level. Another 
 calculation by Singh \cite{bsingh} obtains a strong spin differentiation 
in the electronic 
structure around the Fermi level in the ferromagnetic ground state. 
The majority 
spin Mn-$t_{2g}-$O-$2p$ bands below the Fermi level are 
separated by a gap from the 
Mn-$e_{g}$ derived bands, though there is also a mixing with Tl and O states  
above and below the Fermi level. Thus a small near band edge Fermi surface 
is obtained. For the minority channel instead, a highly dispersive band with 
a strong admixture of Mn, Tl and O is found 
around the Fermi level, leading to a 
metallic minority spin channel \cite{bsingh}.
 
 Mostly due to the absence of evidence for significant doping in the
pyrochlore Mn-O sublattice 
and due to the tendency of Tl to form $6s$ 
conduction bands (unlike in manganese perovskites where the rare-earth levels 
donate charge to the Mn-O bands but 
are otherwise electronically inactive), among 
other differences between the compounds discussed above,   
it has been speculated that a
double exchange mechanism similar to that of perovskites is unlikely
to be effective in Tl$_{2}$Mn$_{2}$O$_{7}$
\cite{5,6,7,5',bsub,bsingh}, 
accounting for the experimental results.   
Based on all these facts, we decided to explore the suitability of 
a different model, 
 to provide an explanation for CMR in pyrochlore Tl$_{2}$Mn$_{2}$O$_{7}$. 
It may be useful to remark here that related pyrochlore compounds have been 
subject of experimental study, and while some (A=Y,Lu) revealed no long 
range magnetic order \cite{sub0}, In$_{2}$Mn$_{2}$O$_{7}$ exhibits a similar 
ferromagnetic transition but the ordered phase remains insulating \cite{6,5'}. 
In this sense, the behaviour of Tl$_{2}$Mn$_{2}$O$_{7}$ is very interesting 
due to its unique characteristics and striking differences with other 
known pyrochlores and CMR compounds.   
 
	 In the following section we will describe the model we employed  
for the calculation of the electronic structure and transport properties  
of Tl$_{2}$Mn$_{2}$O$_{7}$. The intermediate valence (IV) model 
was proposed originally for the
study of Tm compounds \cite{8,9}. 
Basically, a periodic array of mixed valent 
ions, fluctuating between two magnetic 
configurations, would represent the Mn ions appearing in 
Tl$_{2-x}^{3+}$Tl$_{x}^{2+}$Mn$_{2-x}^{4+}$Mn$_{x}^{5+}$O$_{7}$. These would 
be responsible for magnetism.
A conduction band 
would represent the Tl-$6s$ orbitals, which hybridize with the IV ions. 
For simplicity, we do not take into account the O orbitals.
Through hybridization a gap can appear in the density of states. Transport is
due to carriers in the conduction bands. The scattering mechanism
originating from the hybridization with the IV lattice is dependent on  
the magnetic configuration of the lattice, or spin disorder in the material.
 As a result, 
electric conduction and magnetism are intrinsically coupled and CMR results on 
application of a magnetic field.            
In Section~\ref{sec:results} we will make 
a complete presentation and discussion of the results obtained with our 
treatment of the model. A short presentation of our approach to the problem
with some preliminary results has been given before \cite{cocoyoc}. 
In Section~\ref{sec:summary} a summary of our study of 
Tl$_{2}$Mn$_{2}$O$_{7}$ is given.

\section{MODEL AND ANALYTICAL METHOD}
\label{sec:method}    
                      
	The model to be introduced in this section was proposed originally 
for the description of intermediate valence(IV) Tm compounds \cite{8,9}. 
These compounds distinguish themselves from other IV rare earths by 
evidencing a rare sensitivity
to the application of magnetic fields in many physical properties, 
like resistivity, specific heat, thermal expansion, etc.    
The exactly solvable impurity model which 
incorporates the most important feature 
for TmSe, namely valence fluctuations between two magnetic configurations
(corresponding to Tm$^{2+}: 4f^{12}$ and Tm$^{3+}: 4f^{13}$) was shown
\cite{8} to describe most of the peculiar features of the magnetic
properties of paramagnetic intermediate valence Tm compounds.  The
impurity model resistivity exhibits an explicit quadratic dependence
with the magnetization\cite{8}. Such behaviour has also been found in
transport experiments slightly above T$_{c}$ for colossal MR
pyrochlores \cite{5} and manganese perovskites \cite{10}. Employing 
the periodic version of the model \cite{9} the 
phase diagram at T=0 was obtained, 
indicating magnetic ordering. A ferromagnetic 
metallic phase appears for non-stoichiometric samples, while an
antiferromagnetic insulating 
ground state is obtained for stoichiometric ones.   
Calculations of the specific heat and magnetic susceptibility in the
paramagnetic phase treated with the coherent potential approximation (CPA)
\cite{soven} were done and the neutron scattering 
spectrum was studied \cite{9}. Many of these 
results are reminiscent of the behaviour of Tl$_{2}$Mn$_{2}$O$_{7}$ and 
other CMR compounds discussed in the previous section.  In fact, 
for manganese perovskites a similar model
has been considered \cite{11} to propose the possibility of a
metal-insulator transition.

	We will now present the model proposed for Tl$_{2}$Mn$_{2}$O$_{7}$, 
considering that experimental results \cite{7} 
would be compatible with the presence of 
mixed valent Mn$^{4+}$/Mn$^{5+}$ in the form  
Tl$_{2-x}^{3+}$Tl$_{x}^{2+}$Mn$_{2-x}^{4+}$Mn$_{x}^{5+}$O$_{7}$, 
as discussed in 
the previous section. The model \cite{9}
describes a
periodic lattice of intermediate valence (IV) ions, which fluctuate
between two magnetic configurations. To simplify the problem, these 
are associated to single (S=1/2) or
double occupation (S=1) of the ion. It is assumed that the results 
are dependent on the availability 
of two magnetic configurations and not by their detailed structure \cite{8}. 
The IV ions are
 hybridized to a band of conduction
states. The hamiltonian considered is \cite{9}:
\begin{eqnarray}
H \, & \, = \, & \, H_{L} \, + \, H_{c} \, + \, H_{H} \,
\end{eqnarray}
where:
\begin{eqnarray}
H_{L} \, & \, = \, & \, \sum_{j} \left( E_{\uparrow}   \mid j
\uparrow > 
<  j \uparrow \mid   
+  E_{\downarrow}   \mid j \downarrow
> < j \downarrow \mid  \right) \nonumber \\ 
\, &  & \, + \left( E_{+}   \mid j + > 
<  j + \mid  +  E_{-}   \mid j - > < j - \mid \right) \, , 
\nonumber \\ 
H_{c} \, & \, = \, & \,\sum_{k, \sigma} \epsilon_{k,\sigma} \, 
c^{+}_{k, \sigma} c_{k, \sigma}  \, , \nonumber \\
H_{H}  & \, = \, & \sum_{i,j}  V_{i,j} \left(  \mid j + > < j
\uparrow \mid 
c_{i,\uparrow}  +   \mid j - > < j \downarrow \mid 
c_{i,\downarrow} \right) + H.c.  \nonumber 
\end{eqnarray}

H$_{L}$ describes the lattice of IV ions, which for Tl$_{2}$Mn$_{2}$O$_{7}$ 
we would identify with the Mn ions. The S=1/2
magnetic configuration at site j is represented by states $\mid j \sigma >$ 
( $\sigma$ =
$\uparrow, \downarrow$ ) with energies E$_{\sigma}$, split in the
presence of a magnetic field B according to:
\begin{eqnarray}
E_{\uparrow (\downarrow)} \,   \, = \,   \, E - (+) \mu_{0} B . 
\label{eupdown}
\end{eqnarray}
The S=1 magnetic configuration is considered in the highly anisotropic
limit, where the S$_{z}$ = 0 state is projected out of the subspace of
interest as in Refs.[15,16]. This has been done 
for the sake of simplicity, as will 
be discussed shortly.
The S=1 states at site j are then 
represented by $\mid j s >$ (s=+,$-$) and energies E$_{s}$, split by the
magnetic field as:
\begin{eqnarray}
E_{ \pm } \,   \, = \,   \, E + \Delta \mp \mu_{1} B .
\label{emm}
\end{eqnarray}

H$_{c}$ describes the conduction band, which for
Tl$_{2}$Mn$_{2}$O$_{7}$ we would identify with the Tl-$6s$ conduction
band. 

The hybridization term H$_{H}$ describes valence
fluctuations between the two magnetic configurations at one site,  
mediated by the conduction electrons.
For example, promoting a spin up electron into the
conduction band  the IV ion at site j can pass from state $\mid j +>$ to
state $\mid j \uparrow>$. Notice that the highly anisotropic limit
considered inhibits any spin flip scattering processes, 
which would involve transitions between the excluded S$_{z}$ = 0 state of 
the S=1 configuration and the S=1/2 states mediated by conduction electrons 
of opposite spin.
So, the direction of the local spin at each site is conserved, and 
IV ions only hybridize with conduction electrons of parallel spin. This 
will allow for an important simplification in the solution of the 
scattering problem, 
by separating it in two parts associated 
to the spin orientation of the conduction 
electrons. 

The hamiltonian can be rewritten in terms of the following creation
(and the related annihilation) operators for the local orbitals \cite{9} 
(Mn-$3d$ orbitals, in this case) :
\begin{eqnarray}
d^{+}_{j,\uparrow} \, = \, \mid j + > < j \uparrow \mid \; , \nonumber
\\
d^{+}_{j,\downarrow} \, = \, \mid j - > < j \downarrow \mid ,
\end{eqnarray}
for which one has:
\begin{eqnarray}
\left[ d_{i, \uparrow}, d^{+}_{j, \uparrow} \right]_{+} \, = \, 
\delta_{i,j} \left( P_{i,\uparrow} + P_{i,+} \right) \; , 
\nonumber \\
\left[ d_{i, \downarrow}, d^{+}_{j, \downarrow} \right]_{+} \, = \, 
\delta_{i,j} \left( P_{i,\downarrow} + P_{i,-} \right) \, , \nonumber \\
P_{i,+} + P_{i,\uparrow} + P_{i,-} + P_{i,\downarrow} \, = 
\, 1 \, , \nonumber
\end{eqnarray}
where: P$_{j,\alpha}$ = $ \mid j \alpha > < j \alpha \mid $ are
projection operators onto the local magnetic configuration states. 
The local hamiltonian now reads:
\begin{eqnarray}
H_{L} \, & \, = \, & \,  \left( \Delta - \mu_{D} B \right) \sum_{j} 
d^{+}_{j,\uparrow} d_{j,\uparrow} + \left( \Delta + \mu_{D} B \right)
\sum_{j} d^{+}_{j,\downarrow} d_{j,\downarrow} , \nonumber \\
\mu_{D} \, & \,  = \, & \, \mu_{1} - \mu_{0}. 
\end{eqnarray}

	Now advantage will be taken of the type of hybridization present, 
resulting from the highly anisotropic limit chosen 
for the S=1 states.  
 Given a certain
configuration for the occupation of the local orbitals at all sites 
by spin up or down electrons, the spin up conduction electrons will
hybridize only with those IV ions occupied by spin up electrons (i.e. in $
\uparrow$ or $+$ local states). One can simulate this by including a
very high local correlation energy (U~$\rightarrow~\infty$) to be
paid in the event of mixing the conduction electron with ions 
occupied by opposite spin
electrons. Concretely, we take \cite{9}:
\begin{eqnarray}
H \,& \,  = \, & \, H_{\uparrow} + H_{\downarrow} , \\
H_{\uparrow} \, & \,  = \, & \, \left( \Delta - \mu_{D} B \right)
\sum_{j 
\in \uparrow} 
d^{+}_{j,\uparrow} d_{j,\uparrow} \nonumber \\ \, & \, & \, 
 + U \sum_{j \in \downarrow} 
d^{+}_{j,\uparrow} d_{j,\uparrow} + \sum_{k} \epsilon _{k,\uparrow} 
c^{+}_{k,\uparrow} c_{k,\uparrow} \nonumber \\
\, & \, & \, + \sum_{i,j} \left( V_{i,j} d^{+}_{j,\uparrow} c_{i,\uparrow} +
H.c. \right). \nonumber
\end{eqnarray}
H$_{\downarrow}$ is analogous to H$_{\uparrow}$, one having only 
to reverse the sign of the magnetic field and the spin orientation. 
In the following, we neglect the
splitting of the conduction band energies in the presence of the magnetic
field B ($\epsilon _{k,\sigma} \equiv \epsilon _{k}$) , 
and take for the hybridization: V$_{i,j}$ = V~$\delta_{i,j}$.  

	Given a certain configuration of spin orientations distributed
among the sites, one can now solve two separate problems described  
by H$_{\uparrow}$ and H$_{\downarrow}$, respectively. Each corresponds to the 
problem of a band of conduction electrons of a given spin orientation, 
hybridized with a binary alloy characterized 
by the spin orientation of the electrons occupying the local orbitals. 
These problems we solve with a  
generalization of the CPA approximation, as in Ref.[16]. 
A similar treatment was used by Sakai et al \cite{sakai}.  
 We introduce an effective 
diagonal self-energy for the local orbitals, $\Sigma_{(d) \sigma}(\omega)$
for the H$_{\sigma}$ alloy problem, through which on average
translational symmetry is restored. We will now sketch the solution of the 
spin up problem, being straightforward the extension to the  
spin down case \cite{9}. 

For spin up, the effective Hamiltonian reads:
\begin{eqnarray}
H_{eff, \uparrow}(\omega) \, \,  = \,  \, \left( \Delta - \mu_{D} B + 
\Sigma_{(d) \uparrow}(\omega) \right)  \sum_{j}      
d^{+}_{j,\uparrow} d_{j,\uparrow} \nonumber \\    
  + \,  \sum_{k} \, \epsilon _{k} 
c^{+}_{k,\uparrow} c_{k,\uparrow} 
\, + \, \sum_{j} \left( V d^{+}_{j,\uparrow} c_{j,\uparrow} +
H.c. \right). \nonumber
\end{eqnarray} 
In terms of the effective Hamiltonian, the effective pro\-pa\-gator 
in Fourier space can be calculated (a 2X2 matrix here, 
with components associated to the c and d-spin up bands):
\begin{eqnarray}
G_{eff,\uparrow}^{k} (\omega)\, & \,  = \, & \, \frac{1}{\left( \omega -  
H_{eff,\uparrow}^{k}(\omega) \right)} \, .
\end{eqnarray} 
In CPA the effective propagator is taken equal to the ensemble 
average of the propagator determined by H$_{\uparrow}$. The 
average local Green's functions obtained can be expressed:
\begin{eqnarray} 
 <<c_{j,\uparrow}, c^{+}_{j,\uparrow}>>(\omega) \, & = & \, 
\frac{1}{N}  \,
\sum_{k} \, \frac{1}{\omega - \epsilon_{k} - \frac{V^{2}}{\left(\omega - 
\sigma_{(d)\uparrow}(\omega) \right) }} \,   ,    \nonumber
 \\ 
 <<d_{j,\uparrow}, d^{+}_{j,\uparrow}>> (\omega) \, & = & \,   
\frac{1}{\left(\omega - 
\sigma_{(d)\uparrow}(\omega) \right) } \, \nonumber \\   + \, \, \, 
\frac{V^{2}}{\left(\omega - 
\sigma_{(d)\uparrow}(\omega) \right)^{2} } & \, &
<<c_{j,\uparrow}, c^{+}_{j,\uparrow}>>(\omega) \, \, ,
\label{fgreen}
\end{eqnarray} 
where:
\begin{eqnarray}
\sigma_{(d)\uparrow}(\omega) \, & \, = \, & \, \Delta - \mu_{D} B + 
\Sigma_{(d)\uparrow} (\omega) \, .
\label{sigmadup}
\end{eqnarray} 

The CPA equations obtained to determine
self-consistently the effective spin up and down self-energies  
relate them directly to the local Green's functions for the IV ions 
\cite{9}:
\begin{eqnarray}
\Sigma_{(d) \uparrow}(\omega) \, = \, \frac{p - 1}{<< d_{j,\uparrow}, 
d^{+}_{j,\uparrow} >>(\omega) } \, , \nonumber \\
\Sigma_{(d) \downarrow}(\omega) \, = \, \frac{- p}{<< d_{j,\downarrow}, 
d^{+}_{j,\downarrow} >>(\omega) } .
\label{cpa}
\end{eqnarray}
Here p denotes the concentration of spin up sites, 
including single and double occupation of the ion by spin up electrons: 
p $= < P_{i,\uparrow} + P_{i,+} >$. Due to the 
correlations included, for the densities of states hold:
\begin{eqnarray}
\int_{-\infty}^{\infty} d\omega  \rho_{c,\sigma}(\omega) \, & \, = \, & \, 
1 \, , \nonumber
\\ \int_{-\infty}^{\infty} d\omega  \rho_{d,\uparrow}(\omega) \, & \, = 
\, & \, p \, ,  \;
\int_{-\infty}^{\infty} d\omega  \rho_{d,\downarrow}(\omega) \,  \, =
\,  \, 1 - p.
\label{sumrules}
\end{eqnarray} 

 Given a concentration p, which is determined by the magnetization, 
we solve the CPA equations (spin up and down)  
self-consistently together with the total number of
particles equation, through which the chemical potential is determined.
    
	To take into account the effect of temperature 
on the magnetization in a simple form and describe
qualitatively the experimental data in pyrochlores 
\cite{5,7,5'}, we used a simple Weiss molecular field approximation 
to obtain the magnetization at each 
temperature. A similar approach was adopted  
before for calculations of magnetoresistivity in Ce-compounds \cite{Coqblin}. 
For a ferromagnet 
with critical temperature T$_{c}$, saturation magnetization M$_{sat}$ and 
local magnetic 
moments of magnitude $\mu$ which can align with an external magnetic field B 
one has: 
\begin{eqnarray}
\frac{M}{M_{sat}} \, = \, tanh \left( \frac{\mu B}{k_{B} T} + \frac{T_{c} M }
{T M_{sat}} \right) .                  
\label{mag}
\end{eqnarray} 
Concerning the application of this approximation for the magnetization 
to our model for Tl$_{2}$Mn$_{2}$O$_{7}$, we take the local magnetic moment 
$\mu$ coincident with the experimental value for M$_{sat}$, roughly 
3 $\mu_{B}$ per Mn ion \cite{5'}. This is an 
intermediate value between the local moments for
Mn$^{5+}$, of 3.87 $\mu_{B}$, and Mn$^{4+}$, of 2.83 $\mu_{B}$.

Once obtained the magnetization from Eq.~(\ref{mag}), we determine the   
concentration p of ions occupied by spin up electrons using the relation 
between them:
\begin{eqnarray}
M \, = \, [ p - (1-p) ] \mu   \, .                 
\label{mp}
\end{eqnarray}               
This simple equation expresses the 
intrinsic link between the magnetic order measured by
the magnetization, and the transport properties which are determined by the 
CPA self-energies obtained solving the 
alloy problems corresponding to a concentration p of 
spin up sites in the sample,
as will be shown next. 

 We now consider the determination of transport properties.    
 Using the Kubo formula it has been proved before
\cite{velicky} 
that no vertex corrections to the electrical conductivity are 
obtained in CPA for a one band model with diagonal disorder, due to
the short range of the atomic scattering potentials.  
For the same reason mainly, no vertex corrections to the conductivity 
were obtained 
by Brouers et al \cite{brouers} with a CPA treatment of an 
s-d model for disordered noble- and transition-metal alloys. 
Their analysis also applies to our problem.    
Furthermore, in the absence of a direct hopping term between local
orbitals only the conduction band will contribute to the conductivity in our
case \cite{brouers}.

 As in Refs.[23,24,21] we can
obtain the conductivity through Boltzmann equation in the relaxation
time approximation as:
\begin{eqnarray}
\sigma \, & \,  = \, & \, \sigma_{c,\uparrow} + \sigma_{c,\downarrow}
, \nonumber
\\  \sigma_{c,\uparrow} \, & \, = \, & \, n_{c} e^{2} \int 
d\omega \left( - \frac{
\partial f (\omega) }{\partial \omega} \right)
\tau^{k}_{c,\uparrow}(\omega) 
\Phi(\omega),
\end{eqnarray}
where f is the Fermi distribution, n$_{c}$ the total number of
carriers per unit volume, and the relaxation time for 
spin up conduction electrons is:
\begin{eqnarray}
\tau^{k}_{c,\uparrow}(\omega) \, = \, \frac{\hbar}{2 \mid Im \Sigma^{k}_{c
\uparrow} (\omega) \mid}.
\label{tau}
\end{eqnarray}  
From Eq.~(\ref{fgreen}) 
the self-energy for
conduction electrons is related to the effective medium CPA
self-energy through: 
\begin{eqnarray}
\Sigma^{k}_{c \uparrow}(\omega) \, & \,  = \, & \, \frac{V^{2}}{\omega
- \sigma_{(d)\uparrow}(\omega)} \, ,
\label{sigmac}
\end{eqnarray}
and is actually k-independent.
                  
$ \Phi \, = \, \frac{1}{N} \sum_{k} v^{2}_{c}(\epsilon_{k})
\delta(\omega - \epsilon_{k}) $, where $v_{c}(\epsilon_{k})$ is the
  conduction electron velocity.
The extension of these formulae for spin down is straightforward. 

Considering temperatures much lower than the Fermi temperature, one
can approximate: $\Phi (\omega) \sim v_{F}^{2} \rho_{c}^{(0)}(\omega)$,
 being $ v_{F} $ the Fermi velocity. The 
results presented here were obtained assuming
for simplicity a semielliptic density of states for the bare 
conduction band $\rho_{c}^{0}(\omega)$. In this case, as can be seen 
from Eq.(\ref{fgreen}),  
the local Green's function for conduction electrons also will have a  
semielliptic form( with the energies shifted by $\Sigma^{k}_{c 
\sigma}(\omega)$).

\section{RESULTS AND DISCUSSION}
\label{sec:results}
 
 We now present the results  obtained for 
 this model with parameters: W=6eV for the semielliptic bare 
conduction half-bandwidth (centering the conduction band at the
origin), $E=0$, $\Delta$ = $-$4.8 eV, V = 0.6eV. 
To reproduce qualitatively
the experimental magnetization data \cite{5,7,5'} we take T$_{c}$ as 142K 
and a saturation value for the magnetization 
of 3$\mu_{B}$ per Mn ion (like that of free Mn$^{4+}$ ions) in the Weiss
 molecular field 
approximation, as well as $\mu_{D}= 1 \mu_{B}$. In    	 
Fig.~\ref{fig:mt} we plot the temperature dependence of the
 magnetization obtained  
for different values of the magnetic field.   

In Fig.~\ref{fig:dered}.a we show the hybridization gap obtained in the 
CPA spin up densities of states at T=0 with those 
parameters. No gap occurs in the spin down c-density of states, 
as there is strictly 
no spin-down d-density of states to hybridize with in the ordered state (p=1),
 due to the sum rule of Eq.~(\ref{sumrules}).   
  In Fig.~\ref{fig:dered}.b we observe that 
at temperatures above T$_{c}$ there 
still are hybridization gaps present for spin up bands, 
and similar gaps have been opened for the spin down c and d-bands 
(or at least 
pseudogaps, as is the case for spin down bands at the higher 
magnetic fields considered: B=4T,8T).
 
The temperature 
dependence of the chemical potential is plotted in Fig.~\ref{fig:mut}
for 
various values of total number of particles: 
below, through and above the 
hybridization 
gap. Notice that the chemical potential is only sensible to the magnetic 
field when it falls inside the gap, outside which the temperature 
dependence is also very smooth.

In Fig.~\ref{fig:resred}.a we show the resistivity curves obtained for 
three different values of filling, in the absence of magnetic field.
The results can be qualitatively understood in the following way. 
At T=0 the  resistivity is zero 
 due to the absence
of spin disorder, hence of scattering as can be seen quickly from
Eqs.~(\ref{cpa}), (\ref{tau}), 
(\ref{sigmac}) and (\ref{sigmadup}) with $p=1$. 
At finite temperatures 
the "impurity approximation" \cite{8} is useful to explain the 
results obtained: 
a given spin component of resistivity 
is proportional to the 
conduction electron relaxation 
time for that spin direction at the Fermi level, which in 
turn is  proportional 
to the d-density of states of that spin direction at the same energy.
 At zero temperature 
there is no spin down resistivity due to the absence of spin down
d-density of states, as obtained from Eq.~(\ref{sumrules}). 
As temperature increases, spin disorder appears and 
the d-down density of states 
becomes finite, hence resistivity increases. At T$_{c}$ paramagnetism 
 sets in, and
the d-densities of states for both spins tend to become equal, hence
the total 
resistivity 
takes a value about half of theirs. The small negative slope of
resistivity at B=0 above T$_{c}$,
can be adscribed  to the decrease of the d-density of states 
at the Fermi level with temperature. The  
absolute values of resistivity depending on filling are distributed  
  corresponding to the respective values of d-density of 
states at the Fermi level. 
In cases when the Fermi level falls inside the hybridization 
gap our resistivity 
results seem not very reliable, 
in accordance with the limitations of validity 
of the Boltzmann treatment for transport \cite{peierls}.    

In Fig.~\ref{fig:resred}.b we depict the resistivity curves obtained 
for different 
magnetic fields at a filling corresponding to having placed the Fermi level 
slightly above the gap, n=1.085. This is 
the situation of interest in order to explain the transport properties of
 Tl$_{2}$Mn$_{2}$O$_{7}$, as can be seen from the strong 
resemblance of these curves to 
the experimental data around T$_{c}$ \cite{5,6,7,5'}. An order of magnitude
estimate for 
our resistivity results for n= 1.085 is compatible with the values 
experimentally found (see Fig.~\ref{fig:resred}.b). In any case the 
absolute value of resistivity we obtain depends on the total number of
particles, as shown in Fig.~\ref{fig:resred}.a. 

Fig.~\ref{fig:resred}.c  
shows  the 
two components of the resistivity, one for each spin direction, 
at  B= 1 Tesla and two fillings of 
Fig.~\ref{fig:resred}.a. For a low total number of particles such 
that the Fermi level falls 
below the hybridization gap (n=0.15, here), the results 
can be qualitatively understood employing 
the impurity approximation. Here, 
at T=0 the spin up resistivity is zero 
 due to the absence
of spin disorder, resulting in a null self-energy. Also, the d-down 
density of states is zero 
(note the sum rule of Eq.~(\ref{sumrules})), hence 
there is no spin down resistivity at T=0.  
Increasing temperature both resistivities build up along with spin disorder,
 dominating the spin up one due to a higher d-up 
density of states at the Fermi level.
Above T$_{c}$, one enters the 
paramagnetic phase and both resistivities tend to be very 
similar (equal for B=0). For a higher filling, e.g. the Fermi level 
slightly above the gap, like n=1.085, 
the spin components of resistivity appear inverted. Here  
hybridization gap  effects on the CPA self-energy
become relevant. At T=0, spin order again causes zero resistivity. 
But increasing 
temperature, the presence of the spin up hybridization gap near the Fermi
level produces an important increase in the real part of the CPA-self
energy $\Sigma_{(d) \uparrow}$ for those energies. 
At temperatures below T$_{c}$, the absence
of a hybridization gap in the spin down density of states instead, is
reflected in a smoother behaviour (and lower values) of the real part of
the CPA-self energy $\Sigma_{(d) \downarrow}$. As a result of this 
the spin up relaxation time (inversely proportional to the real part
of the alluded CPA-self energy) and resistivity become smaller
than their spin down counterparts. This effect of split band behaviour
on the real part of the CPA-self energy is well known from the general
CPA theory \cite{vke}. For n=0.15, as the Fermi level falls in a region with 
enhanced d-up density of states, a compensation of the effects  
of the real and imaginary parts of $\Sigma_{(d) \uparrow}$ on the 
relaxation time occurs, and the resistivity results can be explained in terms
of the "impurity approximation" as done above. 
The absolute values of 
resistivity obtained for n=1.085 are lower than those for
n=0.15, 
in accordance with general lower values of 
d-density of states. For cases of intermediate 
filling, with the Fermi level inside the gap, we find that the spin 
components of resistivity interpolate between the two regimes 
just described.   

 In Fig.~\ref{fig:mrred} we plot the 
magnetoresistance results obtained 
employing the same parameters as above for the model. Here 
our results exhibit the difference in value between the
magnetoresistivity maxima around
T$_{c}$ for lower (measured by $MR_{14}$ here) and higher
magnetic fields (given by $MR_{48}$) which is present in the 
experimental data \cite{7}. We also reproduce
 the crossing of the magnetoresistivity curves 
for lower and higher magnetic fields which is   
found at temperatures above T$_{c}$. Our magnetoresistivity  
peak values for n= 1.085 are approximately a factor  
 three smaller than those which can be   
obtained from the data of Ref.[9].  
The features mentioned here as present in the experimental 
magnetoresistivity data, 
were determined performing a scan 
of the resistivity curves of Ref.[9] and afterwards processing those data. 
 Magnetoresistivity results obtained for fillings 
such that the Fermi level is placed inside either band, 
below and above the 
hybridization gap, are quite independent of the precise value of filling, 
which is not the case when the Fermi level falls inside the gap region. 
There, our results would indicate that higher values of magnetoresistance 
can be attained and shifts of MR peaks appear. In all, our description 
of the main features exhibited by transport measurements in 
Tl$_{2}$Mn$_{2}$O$_{7}$ around T$_{c}$ 
is quite remarkable, 
considering the simplifications adopted in our treatment. 

We will now refer to the implications for Hall transport experiments. 
In our case, two kinds of carriers are contributing to the transport, namely 
conduction electrons of either spin direction. The 
ordinary Hall coefficient for the 
system, $R_{H}$, 
in terms of those for each type of carrier, $R_{\sigma} \, 
(\sigma=\uparrow,\downarrow), $ takes the form \cite{peierls}:
\begin{eqnarray}
R_{H}(\omega) \, & \,  = \, & \, \frac{R_{\uparrow} \, \sigma^{2}_{\uparrow} +
R_{\downarrow} \, \sigma^{2}_{\downarrow} } { \sigma^{2} }  \, .
\label{rheqn}
\end{eqnarray}                                         
In the 
ferromagnetic phase, 
additional "extraordinary" or "spin Hall" terms in $R_{H}$, 
accounting for coupling between orbital motion of the carriers and 
the magnetization, may become important \cite{rebook}. 

The reports on Hall experiments for Tl$_{2}$Mn$_{2}$O$_{7}$ \cite{5} indicate 
the presence of electron-like carriers, 
about 0.001 conduction electrons per formula 
unit above T$_{c}$. In the ferromagnetic phase,  
the Hall coefficient is reported \cite{5} to be field independent 
(for magnetic fields up to 6 Tesla), no anomalous Hall signal is observed, 
and the number of carriers at low temperatures is about three times 
the value above T$_{c}$. 
We will interpret these facts on the basis of the results obtained 
with our model. The absence of anomalous Hall contributions below T$_{c}$ 
would 
indicate that the 
ordinary Hall coefficient should suffice to explain the data.   
We shall assume a simple free-electron form for the Hall coefficient of each 
carrier species, namely $R_{\sigma} = 1/ecn_{c,\sigma}$, where $n_{c,\sigma}$ 
denotes the number of conduction electrons with spin $\sigma$ 
contributing to transport. Consider the model with the parameters used above,
and a filling such 
as n=1.085, i.e. the Fermi energy placed slightly above the gap 
in the density of states.  
At very low temperatures, the resistivities for spin up and spin down are both 
negligible.
In such case, from Eq.~(\ref{rheqn})  
the Hall coefficient 
would result: $R_{H} \, \sim \, (R_{\uparrow} + R_{\downarrow})/4 $. 
In the paramagnetic 
phase above T$_{c}$, again both conductivities tend to be similar, 
though being finite now, and the same formal expression for $R_{H}$ holds. 
Nevertheless 
the difference in $R_{H}$ above and 
below T$_{c}$ can be explained through the necessary
change in $R_{\downarrow}$. With a gap (larger than $k_{B} T$) present in 
the density of states, and lying the Fermi level (slightly) above the gap, 
it is only carriers in that upper ("conduction") band which are effective 
for transport.                                              
This holds for spin up electrons up to room temperature in our case, the 
resulting spin up Hall coefficient having a negligible temperature dependence. 
For spin down electrons 
instead, there is no gap present in the ferromagnetic phase.
Only at T$_{c}$ has the spin down c-density of states developed a gap, similar 
to the spin up one. Due to this fact, the number of spin down electrons 
effective for transport is reduced above T$_{c}$, resulting in an increase of 
$R_{\downarrow}$ and $R_{H}$ in accordance with 
experimental observations \cite{5,7}.

Meanwhile, it is the total number of conduction electrons (which includes
those below the gap and with our parameters is about 0.15, about ten times
the number of carriers above the gap)
which would represent the real doping (x) into the Mn$^{4+}$ state 
to take into consideration. In this way, the difficulties 
arising from the 
small number of carriers obtained from     
Hall data which are 
mentioned in Refs.[7,9] could be solved. It is interesting to note that 
a similar difficulty regarding the number of carriers detected and those doped 
into the sample had been mentioned by the authors of Ref.[28] when 
presenting their Hall data for TmSe.   
 
	To end this section, we now briefly comment on the transport
results obtained with the impurity version of the model discussed
above \cite{8}. Using parameters for the impurity in accordance with
those employed here for the periodic model, and a filling such that the Fermi 
level falls slightly above the peak of the impurity density of states, we
obtain magnetoresistance results very similar to those 
presented above. Nevertheless, the impurity
picture would be hard to reconcile with the Hall data in 
Tl$_{2}$Mn$_{2}$O$_{7}$ indicating 
a very small number of carriers effective in transport \cite{5,7}.

\section{SUMMARY}
\label{sec:summary}      
                           
Summarizing, in this paper we have presented a complete discussion of the 
magnetotransport properties obtained with the intermediate valence model 
model for the CMR pyrochlore compound  
Tl$_{2}$Mn$_{2}$O$_{7}$ which was proposed in Ref.[17]. 
The model had been used before for the description of 
intermediate valence Tm compounds \cite{8,9}.  
A lattice of intermediate
valence ions fluctuating between two magnetic configurations describes the
Mn ions, which are hybridized with a conduction band which we relate to
Tl. In this model, ferromagnetism originates from the IV lattice, but 
transport is intrinsically coupled to the magnetic configuration of
the lattice (or spin disorder) through the scattering mechanism 
for conduction electrons which is determined by the hybridization. As a result
of this coupling between magnetic order and transport properties,
colossal magnetoresistance is obtained.       
We would like to observe that in spite of the
simplifications adopted in our treatment
of the model and the calculation of magnetotransport properties, 
the qualitative agreement 
between our results and the experimental data available on CMR pyrochlore 
Tl$_{2}$Mn$_{2}$O$_{7}$ is quite remarkable. 

In connection with the model proposed here, it is important to
mention that 
Shimakawa et al \cite{5'} in a recent paper comment that 
the ferromagnetic metallic 
state and the magnetoresistance in pyrochlore Tl$_{2}$Mn$_{2}$O$_{7}$ 
are reminiscent of the 
 s-f interaction in europium chalcogenides and make a suggestion in favour of
 a mechanism for CMR similar to the one we propose \cite{cocoyoc} and 
our calculations substantiate. Europium chalcogenides  
 have some features in common 
with mixed valent Tm compounds, for which
our model has been originally proposed \cite{8,9}. 

The presence of hybridization gaps or pseudogaps in the electronic structure,
such as those considered here, could not only 
solve the problems posed by the Hall data which are mentioned by the
authors of Refs.[7,9]. They should cause observable
effects in other experiments, such as spin polarized tunneling and optical
properties which are interesting to investigate.  
It is interesting to note  
that electronic structure calculations by Singh \cite{bsingh} for the
ferromagnetic phase posess many features in 
common with our proposed band structure.  

Concerning the 
strongly anisotropic approximation adopted 
here for the S=1 magnetic configuration, 
it shouldn't introduce appreciable error for transport in the strongly 
ferromagnetic phase where 
spin flip scattering processes 
are not allowed. The application of a magnetic field 
will also reduce spin flip scattering processes, so we wouldn't expect 
 major deviations from our magnetotransport predictions
 by relaxing this approximation. In any case, in the framework of the model 
and assuming  
Tl$_{2-x}^{3+}$Tl$_{x}^{2+}$Mn$_{2-x}^{4+}$Mn$_{x}^{5+}$O$_{7}$ the
magnetization should be calculated 
considering S=1 and 3/2
magnetic configurations for the IV ions and deriving the
magnetization from the free energy, though we do not expect 
major differences for magnetotransport results. 
 A more accurate treatment of
transport in presence of a hybridization gap in the density of states
would be required to obtain better results for fillings such that the
Fermi level lies in the gap region. Also, a realistic calculation of
the Hall coefficient for this material is a non-trivial and
interesting problem.

\newpage

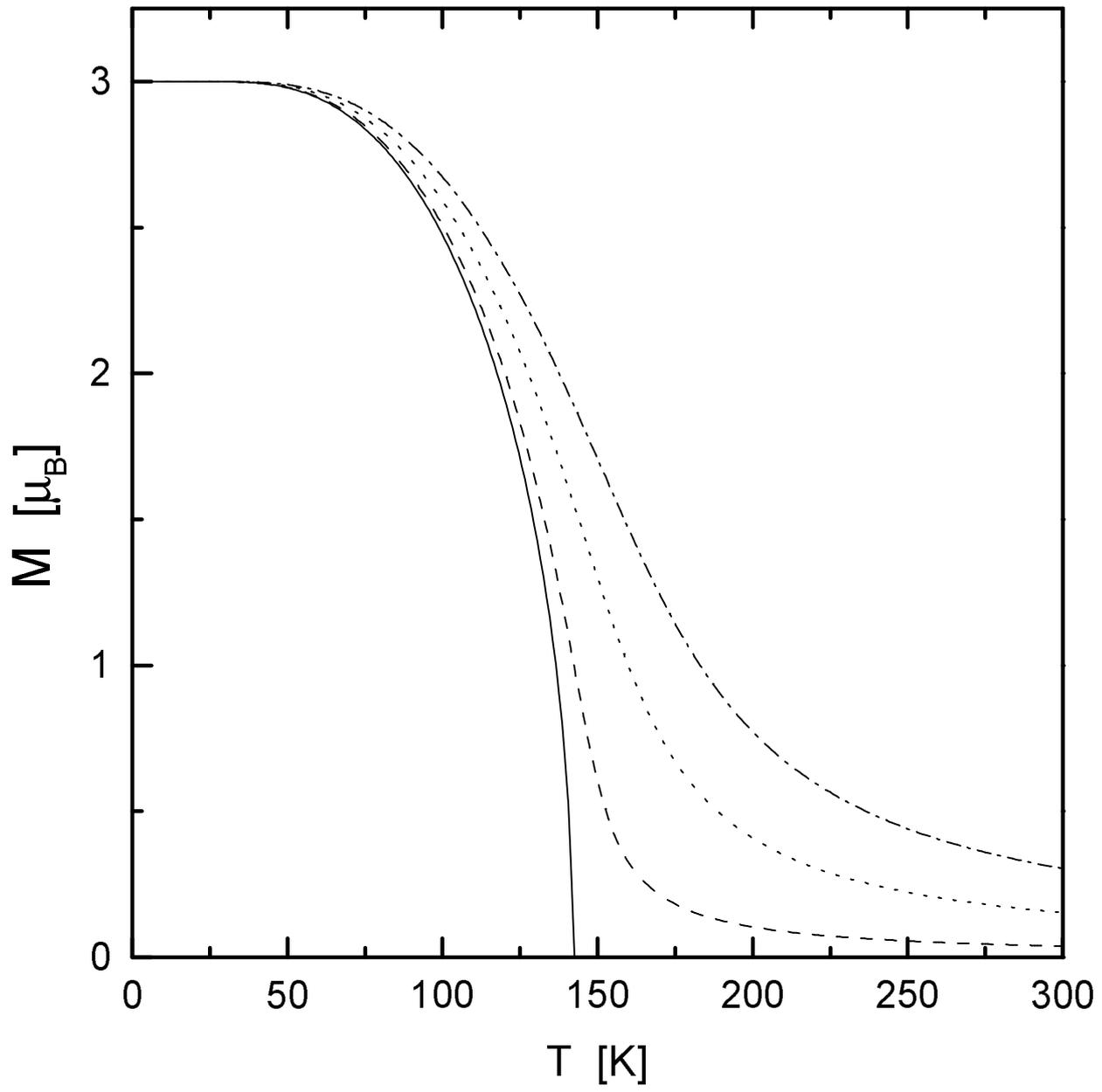
\begin{figure}
\caption{Magnetization as a function of temperature for magnetic
fields: B= 0 (solid line), 1 T (dashed line), 4 T (dotted line), 
8 T (dot-dashed line). Parameters: T$_{c}$= 142~K, M$_{sat}$=
3 $\mu_{B}$, $\mu_{D}$= 1 $\mu_{B}$.}
\label{fig:mt}
\end{figure}

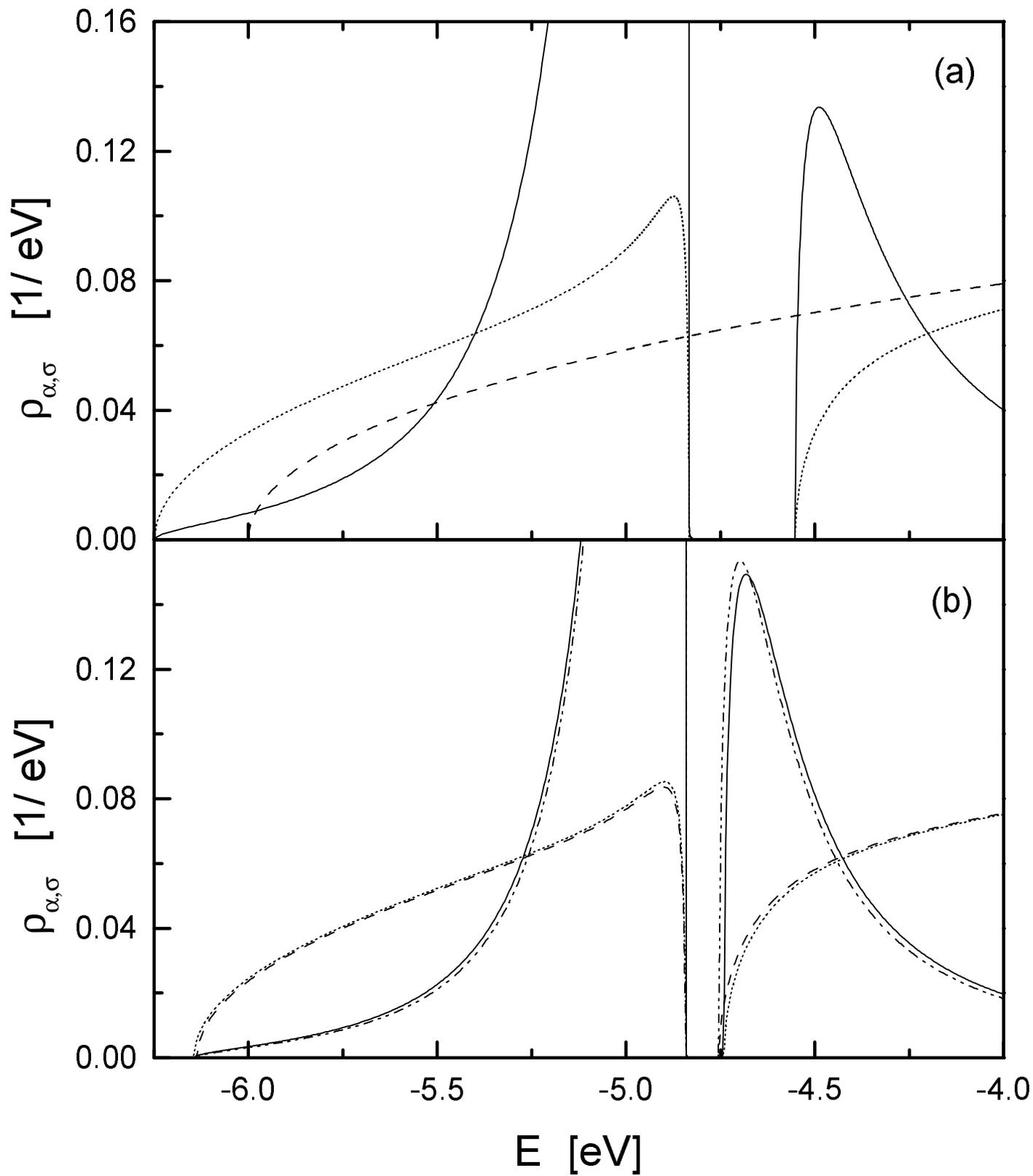
\begin{figure}
\caption{Local densities of states as a
function of energy at B= 1 T and (a) T= 0; (b) T= 200 K. 
 $\rho_{c,\uparrow}$ (dotted line), $\rho_{c,\downarrow}$ (dashed line), 
$\rho_{d,\uparrow}$ (solid line),
$\rho_{d,\downarrow}$ (dot-dashed line).
Parameters: W= 6 eV, 
E=0, $\Delta=$ -4.8 eV, V= 0.6 eV, others as in Fig.1 . }
\label{fig:dered}
\end{figure}

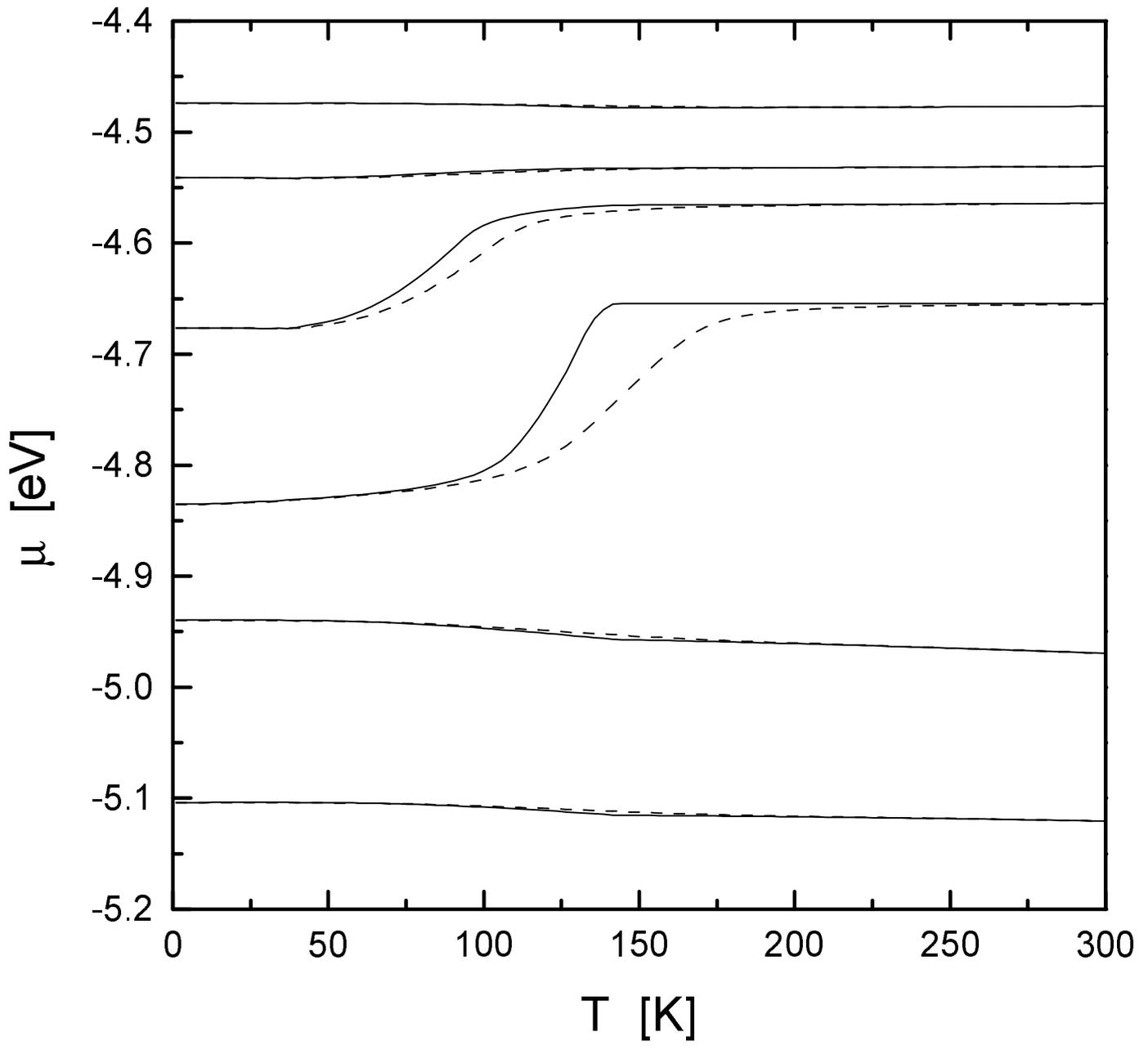
\begin{figure}
\caption{Chemical potential as a function of temperature. 
For various fillings, from top to bottom: n= 1.085, 1.07, 1.06,
1.03, 0.3, 0.15. Full lines: B= 0, dashed lines: B= 8 T.
Parameters as in Fig.2 . }
\label{fig:mut}
\end{figure}

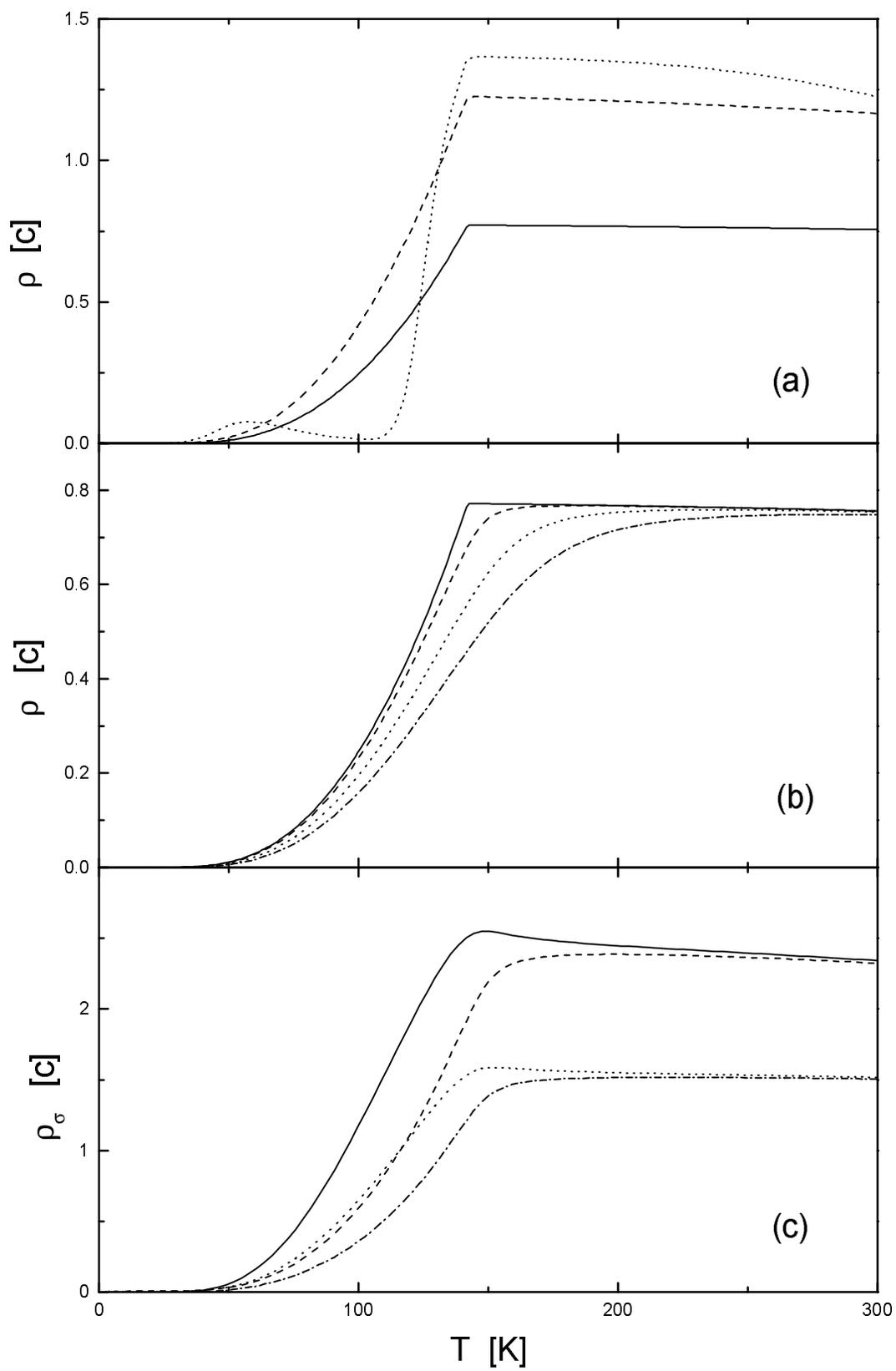
\begin{figure}
\caption{Resistivity as a function of 
temperature. 
(a) Dependence on filling at B=0 : n= 1.085 (solid line); 0.15 (dashed line); 
1.055 (dotted line). (b) Dependence on magnetic field 
at n=1.085 :  B= 0 (solid line); 1 T
(dashed line); 4 T (dotted line); 8 T (dot-dashed line). (c) Spin 
components of resistivity at B= 1 T. At n= 0.15: $\rho_{\uparrow}$ 
(solid line),
$\rho_{\downarrow}$ (dashed line); at n= 1.085: $\rho_{\uparrow}$ (dot-dashed 
line), $\rho_{\downarrow}$ (dotted line).
Parameters as in Fig.2 . 
$c = (eV)^{2}/ [n_{c} e^{2}
v_{F}^2 \hbar]$, e.g. $\, c \sim 0.1 \, \Omega cm$ for $n_{c} \sim
10^{21}/cm^{3}$ and $v_{F} \sim 10^{7}cm/s$.} 
\label{fig:resred}
\end{figure}

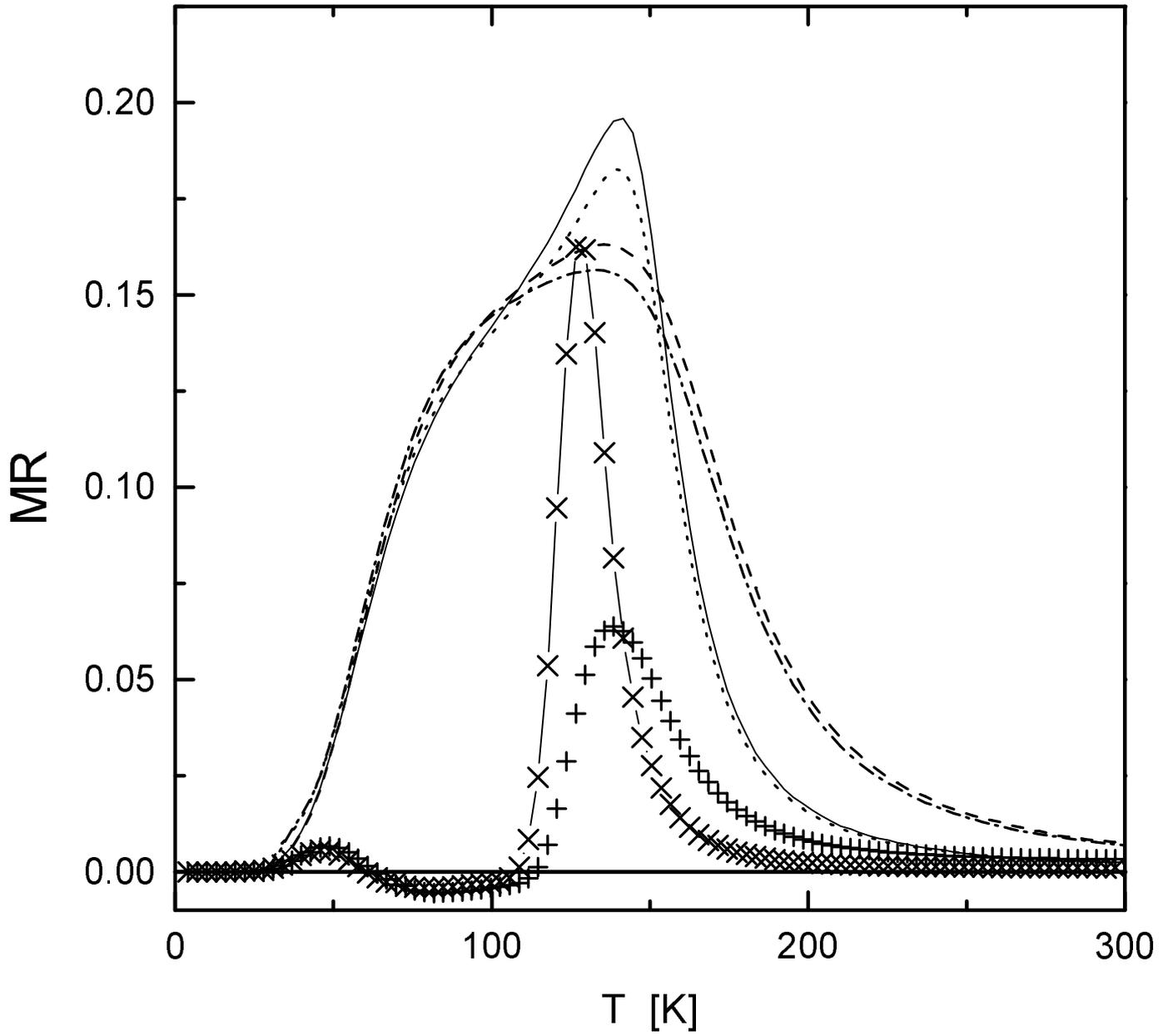
\begin{figure}
\caption{Magnetoresistivity as a function
of temperature. MR$_{14} = \frac{\rho_{1T}(T)
-\rho_{4T}(T)}{\rho_{4T}(T)}$, where $\rho_{B}(T)$ is the resistivity 
at temperature T and magnetic field B, for : n= 1.085 (solid line), 0.15 
(dotted line), 
1.055~(X~crosses). MR$_{48} = \frac{\rho_{4T}(T)
-\rho_{8T}(T)}{\rho_{4T}(T)}$ for : n= 1.085 (dashed line), 0.15 (dot-dashed 
line), 1.055~($+$~crosses). (MR data for n=1.055 plotted 
divided by 10.) 
Parameters as in Fig.2 . }
\label{fig:mrred}
\end{figure}

\end{document}